\documentclass{mem}
\usepackage{natbib}
\usepackage{txfonts}
\usepackage{balance}
\usepackage{graphicx}
\begin{document}

\newcommand{\reason}[2]{ 
\begin{quote}
{\bf #1}: {\em #2}               
\end{quote}
}

\newcommand{\term}[2]{\mbox{$\,^{#1}{\rm #2}$}}

\newcommand{\figjudgeone}{
\begin{figure*}[t!]
\resizebox{\hsize}{!}{\includegraphics[clip=true]{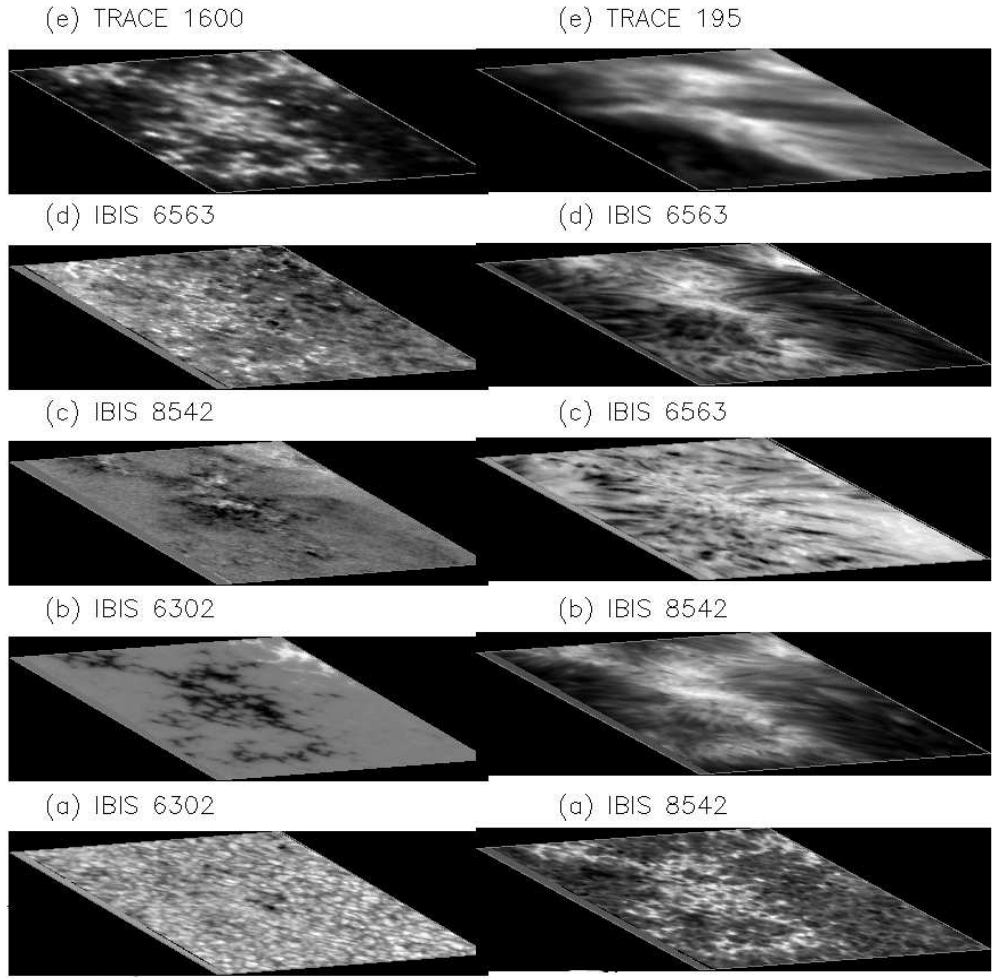}}
\caption{\footnotesize Joint observations of pores from IBIS and TRACE
  from 20 May 2008 (Judge et al., in preparation). The left hand
  panels show (a) the photospheric continuum near 630.2 nm, (b) and
  (c) ``magnetograms'' in the photosphere and chromosphere
  respectively,  (d) far wing of H$\alpha$, (e) 160 nm continuum from
  TRACE.  The right panels show  (a) wing of \ion{Ca}{II} 854.2 nm, (b) core
  of 854.2, (c) and (d) H$\alpha$ 0.8 and 0.1 \AA\ blueward of line
  center, (e) TRACE 19.5 nm coronal loops and footpoints.  The field
  of view is 32 Mm on a side. The fibril structure, requiring $\sim
  1\arcsec$ resolution to see clearly, of both  H$\alpha$ and Ca II
  854.2 nm  chromospheric lines, outlines the  morphology of the base
  of the overlying corona.  }
\label{judge_figone}
\end{figure*}
}

\newcommand{\figjudgetwo}{
\begin{figure*}[t!]
\resizebox{\hsize}{!}{
\includegraphics[clip=true]{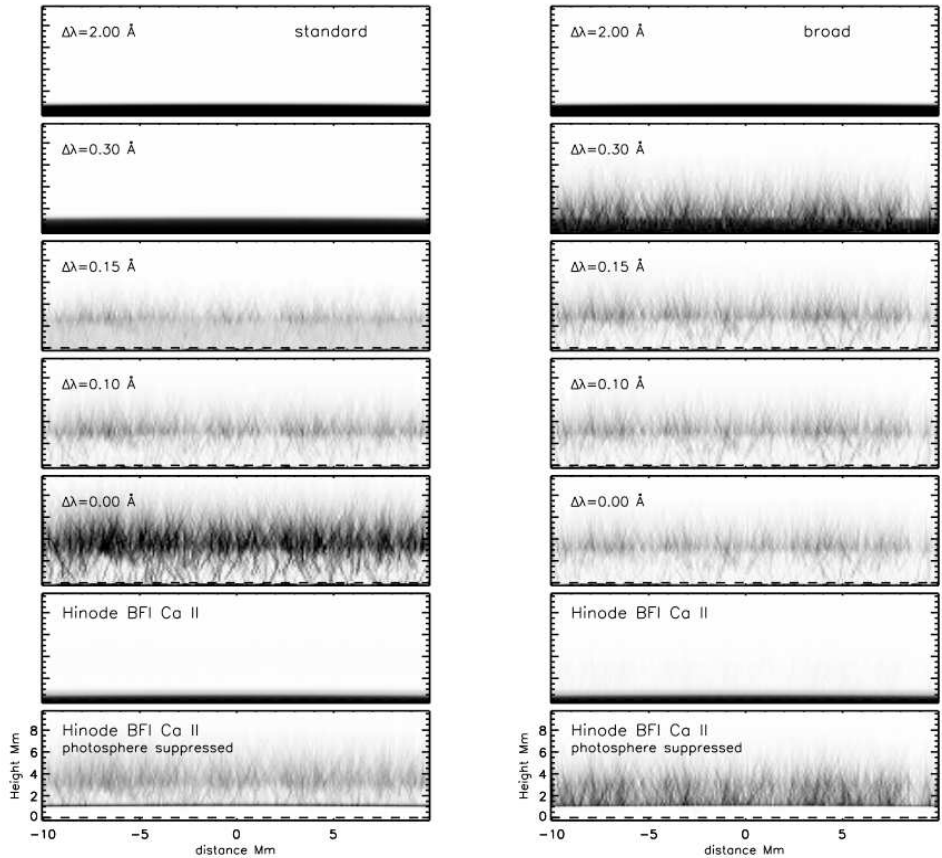}}
\caption{\footnotesize Intensities computed at several monochromatic
  wavelengths and in the Hinode BFI passband are shown as a function
  of position along the limb tangential direction and radial height.
  The color table is reversed for clarity: bright intensity=dark
  images. ``Standard'' spicule conditions were applied in which
  spicules have the same line profiles as the ambient inter-spicule
  medium (left panel), and broad spicular emission lines were
  computed (right panel). Observations are qualitatively similar to
  the ``broad'' calculation.  }
\label{judge_figtwo}
\end{figure*}
}

\newcommand{\figjudgethree}{
\begin{figure*}[t!]
\resizebox{\hsize}{!}{\includegraphics[clip=true]{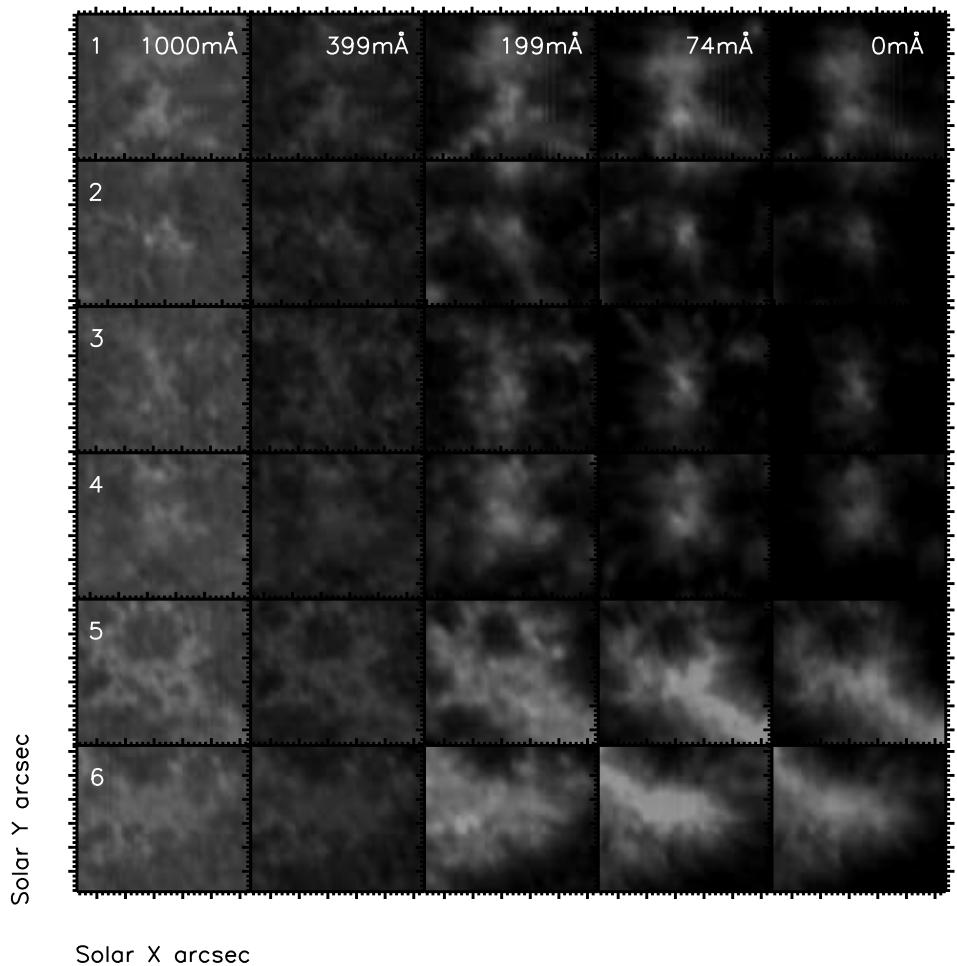}}
\caption{\footnotesize Images constructed from Echelle spectrograph
  observations from June 13 2007, binned over 22.5 m\AA, are shown
  for several different regions of network  emission (labeled
  1-6). The intensity scales are are the same scale for all
  images. Each region is $30\arcsec \times 30\arcsec$ in  size, on
  the order of a typical supergranule cell size. Along each row the
  wavelength varies from wing to line center, thus the rightmost
  images are formed highest in the chromosphere.  }
\label{judge_figthree}
\end{figure*}
}

\title{The chromosphere: gateway to the corona?}
\subtitle{\ldots{}Or the purgatory of solar physics?}
\titlerunning{la cromosfera:il purgatorio del sol?}

\author{P. G. \,Judge}

\offprints{P. G. Judge}

\institute{
High Altitude Observatory, 
National Center for Atmospheric Research\footnote{The National % 
Center for Atmospheric Research is sponsored by the %
National Science Foundation},
P.O.~Box 3000, Boulder CO~80307-3000, USA\\
\email{judge@ucar.edu}
}

\authorrunning{Judge}

\abstract{I argue that one should attempt to understand the solar
  chromosphere not only for its own sake, but also if one is
  interested in the physics of: the corona; astrophysical dynamos;
  space weather; partially ionized plasmas; heliospheric UV
  radiation; the transition region. I outline curious observations
  which I personally find  puzzling and deserving of attention.
   }

\maketitle{}

\section{Introduction}

A cursory glance at the literature reveals that chromospheres are
something of an astronomical backwater. Of $10^6$ abstracts listed in
the ADS (astrophysics section) between 2000 and July 2009, a mere
$4.5\times10^3$ have chromosphere in the abstract, slightly less than 
planetary nebulae and only twice the number of papers of something
entirely invisible, dark matter. Its repugnance presumably 
lies in the difficult regimes of non-LTE and non-ionization
equilibrium, equipartition of magnetic and dynamical energies, 
its awesome fine scale structure, as well as other potential horror
stories involving its poorly understood connections to the photosphere
and corona. 

Why, then, should anyone be interested in it? Here I attempt to show
why we can no longer just ``brush the chromosphere under the carpet'',
(magnetic or figurative) and ignore its importance either in a solar
or plasma physics context. I hope to convince the reader that the
chromosphere deserves to be studied by more than an interesting group
of souls who have, like myself, long since lost their way, and become
hopelessly entangled in one of the most awkward parts of the Sun. The
paper discusses also several problems shamelessly of interest to the
present author, narrated with the help of architypal ``lost soul''
Edgar Allen Poe, whose bicentennial birthday anniversary is 2009.

Readers looking for a respectable review from an
observational perspective can look to \citet{Rutten2006}, a somewhat
historical viewpoint is given by \citet{Judge2006}, and recent
exciting results can be found in papers in the
present volume.

\section{Seven reasons why the chromosphere is important}

In the late 1800s, Hale and Deslandres independently developed and
applied the spectroheliograph to lines of \ion{Ca}{II}, known from
eclipses to be prominent in the solar chromosphere.
\nocite{Hale+Ellerman1904} Hale \& Ellerman's (1904) photographic
spectroheliograms of the disk chromosphere revealed the
``chromospheric network'' at various wavelengths in the \ion{Ca}{II}
$H$ and $K$ lines and in H$\beta$, and showed that enhanced
chromospheric emission occurs in ``clouds'' or ``flocculi'' above
photospheric faculae. In the 1950s, the bright network discovered was
found to be correlated with photospheric velocity and magnetic fields,
associated with supergranular motions identified by
\citep{Simon+Leighton1963, Simon+Leighton1964}. The chromosphere is
the source of variable UV irradiation. Since the visual work of
Secchi in the 1870s, the chromosphere seen in light of H$\alpha$ (the
rose colored line responsible for the chromosphere's name) was known
to contain remarkable and beautiful fine structure (network, fibrils,
spicules). In the 1960s, slit spectra with film and electronic
detectors revealed that the chromosphere is dynamic, supporting wave
motions as well as spicules, dynamic fibrils, surges etc. Thus I come
to the first two reasons to study the chromosphere:  
\reason{1}{We do 
not understand from first principles why the Sun is obliged to 
manifest these phenomena.}   
\reason{2}{Variable UV influences the 
heliosphere, including the Earth's atmosphere.}  
A third simple reason is   
\reason{3}{The Sun is not alone.} 
Chromospheres are present whenever a star 
possesses a convection zone. Beginning in
the 1960s, Olin Wilson began monitoring stellar chromospheric emission
in the cores of the \ion{Ca}{II} resonance lines \citep{Wilson1978}. This
record continues today and serves as the prime database by which
stellar activity cycles, manifestations of dynamo action observed in
the Sun, can be studied
\citep{Baliunas+others1995,Lockwood+others2007}. This is because
magnetic fields on the surfaces of convective stars lead to network
and plage heating, directly reflected in chromospheric emission. More
direct measurements of  magnetic fields for solar-like stars are very
difficult, essentially because there are no magnetic monopoles and the
signals in polarized light cancel almost completely, when integrated
over the stellar surface \footnote{One possible exception is if a 
solar-like star presents an essentially unipolar magnetic hemisphere 
to us, by virtue of being seen pole-on, for example.}. Thus we have
reason number 
\reason{4}{Variable chromospheric emission sheds 
light on dynamo properties across the HR diagram.}   
\figjudgeone
The structure of the solar corona is imprinted already the
chromosphere, something seen clearly when  when chromospheric
fibrils/spicules are resolved.  This chromospheric ``fine structure''
\citep[e.g.][]{Secchi1877,Kiepenheuer1953,Athay1976} is now known to
bear similarities in morphology and dynamics to the overlying corona.
Figure \protect\ref{judge_figone} shows an example of one part of a
small bipolar active region -- the difference in physical conditions
photospheric and coronal images is striking. The morphology of images
obtained  near the H$\alpha$ line core is most simply understood as
reflecting a cool component of the low-$\beta$ hydromagnetic system,
which extends into the overlying corona.  Being dependent on unknown
terms in the energy equation, it is not understood what leads to the
mix of cool and hot material that is represented by such data.  This
aside, the important thing is that \emph{hydromagnetic conditions at 
the coronal base can in principle  be diagnosed using upper 
chromospheric lines}. The low-$\beta$ state  of the upper
chromosphere, inferred simply from the morphology (and dynamics) of
fine structure, implies that  it is in a (nearly) force-free state.
Magnetic field measurements made there lead us to reasons number
\reason{5}{Meaningful force-free boundary conditions can be  used for 
extrapolations into the corona, and } 
\reason{6}{Using the 
magnetic virial theorem \citep{Chandrasekhar1961}, the free energy 
can in principle be determined in the overlying coronal volume.}
Chandrasekhar's theorem applied to volume $V$ bounded by surface $S$
with normal {\bf s} can be written \newcommand\cross\times
\begin{eqnarray} \nonumber
 \int_V {\bf r \cdot [ (\nabla \cross B) \cross B]} dV  = \\ \int_V
 \frac{1}{2} B^2 dV + \int_S [ ({\bf B \cdot r}) {\bf B} - \frac{1}{2}
   B^2 {\bf r}] \cdot d{\bf s}.
  \label{eq:1}
\end{eqnarray}
When ${\bf j = \nabla \cross B=0}$, the LHS vanishes, and the result
follows. Consider a volume of the Sun's atmosphere, in a low-$\beta$
state, bounded by the chromosphere at the base and by force-free
conditions extending into the corona itself. Measurements of the
vector field in $S$ then  suffice to determine the total integrated
magnetic energy, from which the potential energy from the surface may
be subtracted. The application of chromospheric lines in this manner
has not yet been realized, but steps are being taken
\citep[e.g.][]{Judge+others2010}.

Presently unknown chromospheric physics serves to modulate the flow of
mass, momentum, energy into the corona
\citep{Holzer+Hansteen+Leer1997}.  Indeed, it is the mass reservoir
required for ``coronal loop scaling laws''
\citep[e.g.][]{Rosner+Tucker+Vaiana1978, Jordan1992}.  But the
chromosphere also modulates the magnetic field as it emerges into the
corona, in two ways. First, force balance requires that ${\bf |j
  \times B| \rightarrow 0}$ above the $\beta=1$ surface which, outside
of umbrae and very quiet Sun, lies within the chromosphere.  Second,
ion-neutral collisions selectively \emph{dissipate} components of ${\bf
  j}$, ${\bf j_\perp}$, perpendicular to ${\bf B}$, with significant
consequences for the nature and stability of emerged magnetic flux
\citep{Leake+Arber2006,Arber+Haynes+Leake2007,Arber+Botha_Brady2009}.
Thus we are led to reason 
\reason{7}{The chromosphere actively sets 
the  boundary conditions for the corona and its  evolution.} The
ion-neutral collisions dissipate magnetic energy more efficiently when
the magnetic fluctuations have higher frequencies, thus the
chromosphere serves as a low-pass filter for certain kinds of magnetic
fluctuations \citep[e.g.][]{dePontieu+Haerendel1998}, reducing any
high frequency photospheric fluctuations from the spectrum of
fluctuations  surviving to the corona. Further discussion of these
points is found in a review by \citet{Judge2009h} .  

That I arrived at the number seven -- the number of deadly sins --
came as a surprise. The deadly sins were rendered a thing of
heavenly beauty in oil on wood by Hieronymous Bosch in 1485,
inscribed ``Beware, Beware, God Sees''. Perhaps Helios is telling
the author something. Perhaps not. In any case the chromosphere's
secrets made manifest will surely be a heavenly, not hellish,
revelation for solar physics.

\section{Current chromospheric challenges}

The chromosphere is of course interesting in its own right, and much
research is rightly focused on aspects of my ``reason number 1''.
Here I take the editors' advice and author's prerogative to again step
back and look at curious aspects of the magnetized chromosphere --  
why the chromosphere has bright network and plage emission,
why they behave as they do, and how the transition region may fit
into the picture.
I take it as given that the chromosphere is controlled by (sub-)
photospheric dynamics, including convective and wave motions, and by
the emergence and transport of magnetic fields into strong
supergranular downflow lanes, thereby producing the network pattern.  

\subsection{``The Purloined Letter''}

\nocite{Poe1844} 
Poe's (1844) iconoclastic detective story reminds 
me of recent observations of the
solar limb from the \emph{Hinode} spacecraft. How so?  Well in both
cases detectives searched high and low, but 
their search has missed something. In Poe's story, a letter is in plain
sight. In the Sun, the (non-spicule) chromosphere must be present and 
in fact dominant in terms of mass, but is so notably missing
from the data that it is obvious by its absence. \emph{Where is the 
limb chromosphere?}
\figjudgetwo
The BFI instrument on \emph{Hinode}, fed by the Solar Optical Telescope 
\citep[SOT][]{Tsuneta+others2008}, has been used to 
observe the H line of \ion{Ca}{II} using 
a 3 \AA\ wide spectral bandpass ($\equiv 230$ km s$^{-1}$).
At and near the solar limb, such data are
characterized by what appears to be a remarkably dynamic,
spicule-dominated chromosphere at the limb of the Sun
\citep[e.g.][]{DePontieu+others2007}. \emph{Little or no absorption} of
spicule light is seen along their lengths. Judge \& Carlsson (2010,
in preparation) present formal solutions to the transfer equation for
given (ad-hoc) source functions, including an ambient stratified
chromosphere from which spicules originate. Now, some absorption must
be expected because there must be material with significant 
opacity in the \ion{Ca}{II}
lines between the photosphere and corona, that is not within the
spicule population.  Evidence for such material is plentiful and 
contained, for example, in disk
observations of the internetwork chromosphere
\citep[e.g.][]{Lites+Rutten+Kalkofen1993a}.  
Figure \protect\ref{judge_figtwo} shows two calculations from Judge \&
Carlsson's formal solutions, the lowest panels showing observables
computed for the BFI. On the basis of these calculations, 
Judge \& Carlsson argue that the \emph{Hinode}
data require the observed 
spicule emission to be significantly Doppler shifted compared with
the ambient atmosphere, either by turbulent or organized spicular
motions. According to Judge \& Carlsson, at the limb, the 
broad bandpass of the Hinode BFI instrument
preferentially detects the bright, dynamic structures whose line
widths and Doppler shifts are sufficient to avoid the absorption by
the intervening material. The non-spicule chromosphere, the
stratified layer between the photosphere and corona which must be
present, is difficult to see in these data.

The consequences of ``not seeing the Purloined Letter'' are perhaps
significant.  Using standard estimates
\citep[][``VAL'']{Athay+Holzer1982,Vernazza+Avrett+Loeser1981}, 
the total spicule type II mass is only
$10^{-4}$ to $ 10^{-5}$ of the mass of the entire chromosphere-
spicules have a far smaller filling factor and density than the
chromospheric base.  Within a network element, the 
enthalpy flux density is a factor of several lower than the 
radiative flux density of $\ge 2\times10^7$ erg cm$^{-2}$ s$^{-1}$ 
of the network chromosphere
(``guestimated'' from VAL , using losses including Fe~II lines from 
\citealp{Anderson+Athay1989}).  
I conclude that the spicules observed comprise a small 
mass and energy fraction of the chromosphere, the bulk of which 
remains unseen by filter instruments like the BFI on \emph{Hinode}.
They are important, however, in that they present a large areal
interface to the corona which will be mentioned below (section 
\protect\ref{subsec:tr}). They appear to supply large amounts of mass
to the corona \citep{Athay+Holzer1982}, and they reveal Alfv\'enic
fluctuations as they propagate magnetic wave energy into the corona
\citep{DePontieu+others2007}. 

\subsection{``The Cask of Amontillado"}

The network chromosphere exhibits a property which, to the present
author, is a curiosity. The bright network emission remains
geometrically confined to overlie the photospheric magnetic field
concentrations, yet the magnetic field where the chromospheric
emission is bright must expand rapidly with height.
Figure \protect\ref{judge_figtwo} shows very narrow band images
obtained using the Echelle spectrograph at the Vacuum Tower Telescope
by \cite{Rammacher+others2008}.
\figjudgethree
The figure shows that the \ion{Ca}{II} emission remains remarkably confined
locally over the photospheric flux concentrations, yet the emission
comes from regions spanning several pressure scale heights.

Natural explanations exist to explain why both bright photospheric and
transition region plasmas are locally confined to the network
boundaries. However, \citet{Judge+kss2010} note that there is a
potential problem for force balance in the chromosphere, in essence
because bright emission usually requires higher gas pressures. 
Empirical bright network models (VAL-like) have a 
higher gas pressure at each geometric height than cell
interior models; the plasma pressure is expected to exceed the
magnetic pressure at least somewhere in the network chromosphere, and
the magnetic pressure is far higher over the network boundary. The
chromosphere has time to equilibrate pressures from network boundary
to cell interior (several tens of minutes) on the life time of
supergranular structures (30 hours or so). So, horizontal pressure
equilibrium is a reasonable approximation, yet the above data show that 
bright emission is steadfastly confined to regions close to the 
network boundary. How can this be?

A ``resolution'' of this
apparent conundrum is to lower empirical (VAL-like) models down by the
couple of scale heights needed to achieve force balance within the
\emph{photosphere} \citep{Solanki+Steiner+Uitenbroeck1991}. But the
present author considers their success as a real surprise given that
the VAL models (1) are identical in the underlying photosphere, and
(2) were constructed to match spatially averaged intensities in the
much higher chromosphere which do not completely resolve the
network.  There is no physical reason why this drop in height
naturally should lead to force balance in the \emph{chromosphere}- the height
drop is determined entirely by photospheric physics, because the
overlying chromospheric conditions are determined by processes
occurring some 5-9 scale heights higher up. While there is of course
a physical link between these layers, the author is truly amazed that the VAL
stratification simply slid downwards can accommodate chromospheric 
force balance, something that is not guaranteed in such modeling efforts.

\nocite{Poe1846}  
Poe's (1846) short story ``The Cask of Amontillado'' is concerned with \emph{
immurement} -- being imprisoned between walls. \citet{Judge+kss2010}
ask if something is missing in our understanding of the force and
energy balance of the chromosphere, or is the network chromosphere
immured by some forces yet to be identified?

%\subsection{The chromosphere is a low pass magnetic filter}

\subsection{``The Fall of the House of Usher''}
\label{subsec:tr}

Summarizing the 1994 NSO Summer Workshop, Eugene Parker commented on
the ``horror of the solar transition region'', in reference to a
work claiming that it consists of filamentary structure filling tiny
fractions of the available volume 
\citep[e.g.][]{Dere+others1987,Judge+Brekke1994}. The author has a feeling
that the latter paper and perhaps other transition
region (TR) literature fits one of Poe's genres: ``horror fiction''. The TR
has the (un-?) fortunate property of radiating a lot of UV
radiation at wavelengths where (1) the background continuum is
relatively weak (wavelengths below 165 nm), and (2) normal incidence
spectrographs are highly reflective (wavelengths above 115 nm). It is
also ``sandwiched'' between the chromosphere and corona, contains very
little mass, and so tends to be spectacularly responsive to
perturbations both from the underlying chromosphere and the overlying
corona. Being, then, a prime target for interesting observations, the
literature measured by solar-stellar ``transition region'' articles in
the ADS constitutes almost 40 \% of that for the literature
chromosphere which spans 9 scale heights and contains 5 orders of
magnitude more mass.  

The story of explaining the lower TR and its \nocite{Poe1839}  
connection to the chromosphere  perhaps has 
parallels with Poe's ``The House of Usher'', a tale with 
dramatic imagination but with tragically
flawed characters. The house itself and 
its inhabitants, a brother and sister with psychological problems, 
are ultimately finished off by a bolt of lightning. No
flash of brilliance has yet brought the transition region story to an
end, but many imaginative physical pictures have been brought to bear
on the problem.

In the 1970s C. Jordan pointed out that resonance lines of neutral and
ionized helium, formed in the lower TR are factors of several 
too bright compared with other
lines, without the need for models \citep{Jordan1975}.   
\citet{Judge+others1995} showed that a similar problem existed
for Li and Na-like ions, using highly accurate irradiance
data. In a step perhaps towards insanity, one
wonders if the basic assumptions behind such emission measure work are
valid for \emph{any} TR line.  
Models dominated by heat conduction produce too little
emission from the lower TR (below $2\times 10^5$ K) by orders of
magnitude, unless somewhat special and questionable geometries are invoked 
to allow cross-field (ion-dominated) heat conduction to occur 
\citep{Athay1990}. This serious
problem was  evident early \citep{Athay1966} and has been expanded on
by many \citep[e.g.][]{Gabriel1976,Jordan1980b}. Such models
also cannot radiate away the downward conductive flux of $\sim 10^6$ 
erg cm$^{-2}$ s$^{-1}$ \citep[e.g.][]{Jordan1980b,Athay1981}, begging
the question, where did it go? Some semblance of sanity returned
perhaps when 
\nocite{Fontenla+Avrett+Loeser2002} Fontenla \emph{et al.} (2002, and
earlier papers in the series) showed that energy balance can be
achieved through field-aligned (1D) diffusion of neutral hydrogen and
helium atoms.  The neutral atoms diffuse into hot regions, radiate
away much of the coronal energy, and can reproduce the H and He line
intensities.  
But alas, there is a serious and nagging problem 
of the peculiar spatial relationship
between the observed corona, TR and chromosphere as observed at
moderate angular resolution ($2-5\arcsec$) \citep{Feldman1983}.  
Feldman and colleagues have since become convinced that 
the lower TR is thermally disconnected from
the corona \citep[e.g.][and references
therein]{Feldman+Dammasch+Wilhelm2001}.   
Yet \citet{Fontenla+Avrett+Loeser1990} declared that ``The above
[i.e. their] scenario explains why (as noted by Feldman 1983) the
\nocite{Feldman1983} structure of the transition 
region is not clearly related to the structures in the corona''.  
That the debate still rages is evidenced by advocates for  ``cool
loop'' models in which lower TR radiation originates from loops
never reaching coronal temperatures and having negligible conduction 
\citep[][and references therein]{Patsourakos+Gouttebroze+Vourlidas2007}.   

A recent addition to this awful, ill house was
prompted by new data suggesting that neither cool loops nor
field-aligned processes adequately describe magnetic and geometric
properties of the Ly$\alpha$ chromospheric network
\citep{Judge+Centeno2008}. The authors, out of desperation, have
declared the Usher sister dead, perhaps even buried her alive, and
appealed to cross-field diffusive processes to allow cool spicular
material to tap into coronal energy and thus generate the bright
radiation from the lower TR \citep{Judge2008}. Unfortunately, 
if the analogy with
Poe's work follows, this scenario too, is doomed.

\section{Conclusions}

The stories of Poe and his life are not happy ones. My parallels end
on a far more optimistic note. New instrumentation is bringing us ever
closer to understanding how the chromosphere is driven, and its
relationship with the neighboring plasmas. Chromospheric vector
spectropolarimetry is with us thanks to new capabilities at infrared
wavelengths \citep[e.g.][]{Solanki+others2003} and using Fabry-P\'erot
interferometers \citep[e.g.][]{Judge+others2010}. The IRIS SMEX
mission scheduled for a 2012 launch will perhaps be able to lay to
rest the House of Usher with all its problems and build a fresh,
lasting picture of the solar transition region free of its horrors,
with its very rapid spectral scanning capability at UV wavelengths,
and sub arcsecond resolution.

\begin{acknowledgements} The author is very grateful to the organizers
for  the invitation to a most interesting meeting, and for
encouragement to present some provocative ideas about the
chromosphere. I thank those participants who were suitably provoked
for their advice, and Alfred de Wijn for a critical reading of the
paper. I thank Terri for 20 wonderful years of marriage and for her
many faceted wisdom, not least her passion for literature.
\end{acknowledgements}

\end{document}